\documentstyle[aps]{revtex}
\draft
\input epsf
\begin{document}
\twocolumn[\hsize\textwidth\columnwidth\hsize\csname@twocolumnfalse%
\endcsname
\title{Reply to Comment on ``Quantum Force in a Superconductor''}
\author{A. \ V. \ Nikulov}

\address{Institute of Microelectronics Technology and High Purity
Materials, Russian Academy of Sciences, 142432 Chernogolovka, Moscow
District, RUSSIA}

\maketitle
\begin{abstract}
{A reply is made to Comment by D.P.Sheehan in which he takes notice that
the quantum force introduced in my earlier paper A.V.Nikulov, Phys. Rev. B
{\bf 64}, 012505 (2001) challenges to the absolute status of the second law
of thermodynamics. It is shown that the introduction of the quantum force
explains only an experimental challenge to the second law. The observation
of the persistent current at non-zero resistance $R > 0$ is experimental
evidence of violation of the postulate saving the second law in the
beginning of the 20 century. According to this postulate any Brownian
motion is absolutely random whereas the persistent current at $R > 0$ is a
regulated Brownian motion.} \end{abstract}

 ]

\narrowtext

D.P.Sheehan writes in the Comment \cite{Comment} that some statements of
\cite{PRB2001} would seem to pit some of the central predictions of quantum
mechanics against the absolute status of the second law of thermodynamics.
He has paid heed that "the combination of the persistent currents and
voltages in the presence of dissipation, particularly if they are driven by
thermal fluctuations, is troublesome since it would appear to allow for the
performance of work solely at the expense of heat energy under equilibrium
conditions". These remarks are correct.

Indeed the combination of the persistent currents and voltages is the
persistent power (i.e. the dc power in the equilibrium state). It is
written in \cite{PRB2001} that the inhomogeneous superconducting loop with
a nonzero persistent voltage $V_{p.v.} \neq 0$ is an electric circuit in
which the $l_{a}$ segment with higher $T_{c}$ is a power source and the
$l_{b}$ segment with lower $T_{c}$ is a load. The existence of such
persistent power source threatens the second law although its violation can
not be without any external load, when all power induced in the $l_{a}$
segment is dissipated in the  $l_{b}$ segment. Real violation of the second
law will take place when an external load with a resistance $R_{l}$ is
linked up the $l_{a}$ segment and the power $W_{l} =
V_{p.v.}^{2}R_{l}/(R_{l}+R_{b})^{2}$ can be used for an useful work.
$R_{b}$ is the resistance of the $l_{b}$ segment.

If the external load is an electricmotor then violation of the Thomson's
formulation takes place, if it is a resistance at higher temperature then
the Clausius's formulation is broken. The total entropy might be
systematically reduced, contrary to the second law, when the power $W_{l}$
is transformed to an ordered energy, for example when an electricmotor
turns a fly-wheel. In this process the chaotic energy of thermal
fluctuations, i.e. the heat energy $Q$, is transformed in the ordered
energy of the revolving fly-wheel, i.e. the total heat energy is
systematically reduced with the velocity $dQ/dt = -W_{l}$ (for the ideal
case) and the total entropy is reduced with the velocity $dS/dt =
-W_{l}/T$.

In order to solve the found contradictions of \cite{PRB2001} with the
second law of thermodynamics D.P. Sheehan proposes two alternatives: 1) the
theoretical description of quantum force system published in Phys.Rev.B
\cite{PRB2001} is incomplete, or 2) either our tradition understanding of
quantum mechanics is incomplete or the second law is violable. Because of
the predominant belief in the inviolability of the second law and of the
strong internal consistency and experimental support for traditional
quantum mechanics D.P.Sheehan has chosen the (1) alternative and implies
the existence of a compensating counterforce which exactly offsets the
quantum force introduced in \cite{PRB2001}.

Any theoretical description may be incomplete and/or incorrect. And if a
description contradicts to the second law almost everyone decides without
any reflection that it can not be correct. Arthur Eddington wrote in 1948
\cite{Eddingto}:  {\it The second law of thermodynamics holds, I think, the
supreme position among the laws of Nature. If someone points out to you
that your pet theory of the universe is in disagreement with Maxwell's
equations - then so much the worse for Maxwell's equations. If it is found
to be contradicted by observation, well, these experimentalists do bungle
things sometimes. But if your theory is found to be against the second law
of thermodynamics I can give you no hope; there is nothing for it but
collapse in deepest humiliation.} Nevertheless I propose to do not conclude
without a careful investigation of the problem.

Even if we pass a resolution that the theoretical description published in
Phys.Rev.B \cite{PRB2001} is incomplete or incorrect the challenge to the
second law could not be removed. D.P.Sheehan writes that the dissipation of
the $I^{2}R$ power during indefinite time threatens the absolute status of
the second law. It is correct statement. Indeed, any power dissipation
during indefinite time can not exist without a power source with $W_{s} =
I^{2}R$. The existence of such power source is the real challenge to the
second law if the current $I$ is not chaotic. The persistent current is not
chaotic and it is observed at $R > 0$ according to the Little-Parks
experiment \cite{little62}.

Thus the initial challenge to the second law is the Little-Parks experiment
\cite{little62}. According to the universally recognized explanation
\cite{tink75} of this experiment the resistance oscillations are observed
\cite{repeat} because of the oscillations of the persistent current. The
work \cite{PRB2001} explains only why the persistent current can be
observed in a loop \cite{repeat} at non-zero resistance $R > 0$ and zero
Faraday's voltage $d\Phi/dt = 0$. The Little-Parks experiment is not only
experimental evidence of the persistent current at $R > 0$. The persistent
current at $R > 0$ was predicted \cite{Kulik} and observed \cite{IBM1991}
also in normal metal mesoscopic loops. But the Little-Parks experiment is
first and most reliable evidence.

The quantum force $F_{q}$ introduced in \cite{PRB2001} takes the place of
the Faraday's voltage $- (1/c) d\Phi/dt$ which maintains the screening
current in a conventional loop with $R > 0$. The current $j$ is stationary
$dj/dt = 0$ in a conventional loop if the electric force $F_{e} = eE = -
(e/cl) d\Phi/dt$ offsets the dissipation force $F_{e} = eE = -F_{dis} = e
\rho j$. Because the dissipation force should act at $\rho > 0$ and $j \neq
0$ the persistent current can exist only if a force offsets the dissipation
force. Therefore the introduction of the quantum force in \cite{PRB2001} is
well-grounded. D.P.Sheehan assumes the existence of a compensating
counterforce which exactly offsets the quantum force. This compensating
counterforce can be only the dissipation force. In the opposed case the
balance of the forces would be broken. If an other compensating
counterforce exactly offsets the quantum force then we should introduce an
additional force in order to explain the observation of the persistent
current at $R > 0$.

According to \cite{PRB2001} the persistent current $j_{p.c.}$ does not go
out slowly at $R > 0$ and $d\Phi/dt = 0$ since the momentum circulation of
superconducting pairs  $\oint_{l}dl p = \oint_{l}dl (2mv_{s} + (2e/c)A)$
should change at each closing of the superconducting state in the loop from
$(2e/c)\Phi$ to $n2\pi \hbar$ because of the quantization. The velocity of
the momentum change equals a force inducing this change. Therefore the
circulation of the quantum force $\oint_{l}dl F_{q} = 2\pi \hbar (n -
\Phi/\Phi_{0}) \omega _{sw}$, where $\Phi_{0} = \pi \hbar c/e$ is the flux
quantum; $\omega _{sw}$ is the average frequency of switching of the loop
between superconducting states with different connectivity.

The quantum force is not connected somehow with the $T_{c}$-inhomogeneity.
It can act in both inhomogeneous and homogeneous loops. But the persistent
voltage can be observed only in an inhomogeneous loop. The possibility of
the persistent voltage predicted in \cite{PRB2001} is obvious from an
analogy with a conventional inhomogeneous loop. According to the Ohm's law
$\rho j = E = -\bigtriangledown V - (1/c)dA/dt = -\bigtriangledown V -
(1/cl)d\Phi/dt$ the potential difference $V = (<\rho>_{l_{s}} -
<\rho>_{l})l_{s}j$ should be observed on a segment $l_{s}$ of an
inhomogeneous loop at $j \neq 0$ if the average resistivity along this
segment $<\rho>_{l_{s}} = \int_{l_{s}} dl \rho /l_{s}$ differs from the one
along the loop $<\rho>_{l} = \oint_{l} dl \rho /l$. Therefore the existence
of the persistent current implies a possibility of the persistent voltage
which can be observed on segments of an inhomogeneous loop. The value and
sign of this voltage, as well as of the persistent current, should depend
in a periodic way on a magnetic flux $\Phi$ inside the loop:
$V(\Phi/\Phi_{0})$. Such voltage oscillations were measured recently on
segments of an asymmetric Al loop \cite{Dubonos}. This experimental result
corroborate that the quantum force as well as the Faraday's voltage is
distributed uniformly among the loop and can not be localized in any loop
segment  \cite{PRB2001}.

D.P.Sheehan proposes the interesting comparison with the case of
semiconductor diodes, which from time to time have been invoked in possible
scenarios for second law violation \cite{McFee}. Indeed, what difference is
between the loop with the persistent current at $R > 0$ and an conventional
electric circuit with semiconductor diodes. It is well known that any
resistance at nonzero temperature is the power source of the thermally
induced voltage called the Nyquist's (or Johnson's) noise. This phenomenon
was described theoretically by Nyquist \cite{Nyquist} and was observed by
Johnson \cite{Johnson} as long ago as 1928 year. The Nyquist's noise is a
dynamic phenomenon in the equilibrium state at nonzero dissipation. Such
phenomena are called Brownian motion \cite{Feynman}.

The persistent current at $R > 0$ is also a dynamic phenomenon in the
equilibrium state at nonzero dissipation. Therefore it is natural that the
maximum power of the persistent current $(k_{B}T)^{2}/\hbar$ \cite{PRB2001}
is close to the total power of the Nyquist's noise. But there is an
important difference between these two Brownian motions. The power of the
Nyquist's noise is chaotic. In the classical limit $\hbar \omega \ll
k_{B}T$ it is "spread" on all frequencies $\omega $, $W_{Nyq} =
k_{B}T\Delta \omega $ \cite{Kittel}, is proportional to the frequency band
$\Delta \omega $ and  $W_{Nyq} = 0$ at $\Delta \omega = 0$. As opposed to
this the power of the persistent current is not chaotic. It is not equal
zero $0 < W_{p.c.} \leq (k_{B}T)^{2}/\hbar$ at $\omega = 0$. Because of
this difference the persistent current at $R > 0$ is the challenge to the
second law whereas the Nyquist's noise is not the one.

The investigations of the Brownian motion in the beginning of the 20
century were very important for the history of physics. Thanks to these
investigations the atomistic-kinetic ideology has prevailed once and for
all the thermodynamic-energetic one prevailing in 19 century
\cite{Smolucho}. The Brownian motion is clear and monosemantic evidence of
the perpetual motion which was considered as impossible according to the
interpretation of the second law prevailing in 19 century. It ought be
emphasized that the Brownian motion is evidence not only of the perpetual
motion but also of a perpetual driving force because no motion is possible
without a driving force at non-zero friction. Therefore this phenomenon was
interpreted as the challenge to the second law in the beginning of 20
century \cite{Smolucho}.

The second law was rescued in that time by the postulate of absolute
randomness of any Brownian motion and it has become the law of chaos
increase. According to the belief prevailing during 20 century the
perpetual motion exists but it is  random and therefore can not be used for
an useful work. In order any Brownian motion (for example the Nyquist's
noise) can be used it should be put in a order (for example to be
rectified). The possibility of the Nyquist's noise rectification has been
considered in \cite{McFee} referred by D.P.Sheehan and in other works
\cite{Brilluin}. Electric diodes can rectify the Nyquist's noise only if a
temperature difference is in the electric circuit \cite{Berger}. The
rectified Nyquist's noise can be used to perform work. But this is not the
challenge to the second law because the work is extracted owing to a
temperature difference as well as in any other existing heat engine
\cite{Berger}.

The rectification of the Nyquist's noise \cite{McFee,Brilluin} appertains
to the numerous challenges to the second law authors of which consider any
possibility to put in a order the chaotic Brownian motion \cite{Berger}.
The best known one  - the Maxwell's demon was posed by Maxwell in 1867. As
opposed to these known challenges the persistent current at $R > 0$ is
already regulated Brownian motion and therefore can be used to perform work
without any  rectification. This phenomenon is possible since the postulate
of absolute randomness of any Brownian motion is not correct according to
the quantum mechanics. According to the classical mechanics the average
velocity of any Brownian motion equals zero: if spectrum of permitted
states is continuous then for any state with a velocity $v$ a permitted
state with opposite velocity $-v$ and the same probability $P(v^{2})$
exists, therefore $\overline{v} = \sum_{per.st.} vP(v^{2}) + (-v)P(v^{2})
\equiv 0$. But according to the quantum mechanics no all states are
permitted. Therefore the average velocity of some quantum Brownian motion
can be non-zero $\overline{v} \neq 0$.

The introduction of the quantum force in \cite{PRB2001} for the description
of the persistent current at $R > 0$ is reminiscent the introduction of the
Langevin force $F_{Lan}$ for the description of the classical Brownian
motion. The quantum force induced by the fluctuations is the Langevin force
with non-zero average value. The latter is very important. The Langevin
equation $$m\frac{dv}{dt} + \gamma v = F_{Lan} \eqno{(1)}$$ can be used for
the description both of the Brownian motion and the motion of an
automobile. The only qualitative difference consists in the character of
the driving force: the average force driving an automobile should be
non-zero, whereas it was postulated that the Langevin force $F_{Lan}$ is
absolutely random and its average value equals zero $\overline{F_{Lan}} =
0$ in all cases. But if $\overline{F_{Lan}} \neq 0$ then the Brownian
motion does not differ qualitatively from the motion of an automobile and
it can be used in order to drive the automobile in violation of the second
law.

The quantum force induced by the fluctuations as well as any Langevin force
is the balancing counterforce to the dissipation force $F_{dis} = -\gamma
v$: $\overline{F_{Lan}} = - \overline{F_{dis}} = \overline{\gamma v}$. This
balance differs qualitatively from the one of the electrostatic and
"diffisive" forces in the p-n junction of semiconductor diodes considered
in \cite{Comment}. The first balance is dynamic $\overline{v} \neq 0$ and
consequently both $\overline{F_{Lan}}$ and $\overline{F_{dis}}$ perform the
work whereas the second balance is static and the work is not performed. As
it is well known the work performed by the dissipation force
$\overline{F_{dis}}$ increases the entropy. Consequently the work performed
by its counterforce $\overline{F_{Lan}} \neq 0$ decreases the entropy in
violation of the second law.

In summary, D.P.Sheehan raises the correct and important question. Some
statements published in Phys.Rev.B \cite{PRB2001} contradict definitely to
the second law of thermodynamics. But it is not correct to conclude without
careful consideration that the analysis made in \cite{PRB2001} is
incomplete. The main threat for the absolute status of the second law is
the experimental observation of the nonchaotic Brownian motion. There is
repeated the situation which was in the beginning of 20 century when the
classical  Brownian motion was interpreted as the challenge to the second
law. The quantum force is introduced \cite{PRB2001} in order to explain the
observation of the persistent current at $R > 0$ just as the Langevin force
was introduced in order to explain the observation of the motion of the
Brownian particles under equilibrium conditions. In order to save the
second law from the new appearing threat an alternative (to \cite{PRB2001})
explanation of the persistent current at $R > 0$ should be proposed.

\end{document}